\documentclass[conference]{IEEEtran}
\usepackage{graphicx}
\usepackage{amsmath}
\usepackage{array}
\newcommand{\PreserveBackslash}[1]{\let\temp=\\#1\let\\=\temp}
\newcolumntype{C}[1]{>{\PreserveBackslash\centering}p{#1}}
\newcolumntype{R}[1]{>{\PreserveBackslash\raggedleft}p{#1}}
\newcolumntype{L}[1]{>{\PreserveBackslash\raggedright}p{#1}}
\usepackage[usenames]{color}
\usepackage{colortbl,booktabs}
\usepackage{tabularx}
\usepackage{multicol}
\usepackage{booktabs}
\usepackage[Symbol]{upgreek}
\usepackage{subfigure}
\usepackage{bm}
\usepackage{cite}
\usepackage{balance}
\usepackage[boxed]{algorithm2e}

\usepackage{threeparttable}
\usepackage{amsmath}
\usepackage{dcolumn}
\usepackage{multirow}
\usepackage{booktabs}
\usepackage{amsfonts}
\newcolumntype{d}[1]{D{.}{.}{#1}}

\def \qed {\hfill \vrule height6pt width 6pt depth 0pt}

\begin{document}

\bibliographystyle{IEEEtran} 
\title{Matrix Inversion-Less Signal Detection Using SOR Method for Uplink Large-Scale MIMO Systems}

\author{Xinyu Gao, Linglong Dai, Yuting Hu, Zhongxu Wang, and Zhaocheng Wang}
\author{
\IEEEauthorblockN{Xinyu Gao, Linglong Dai, Yuting Hu, Zhongxu Wang, and Zhaocheng Wang}
\IEEEauthorblockA{Tsinghua National Laboratory for Information Science and Technology (TNList),\\
Department of Electronic Engineering, Tsinghua
University, Beijing 100084,
China\\
E-mail: daill@tsinghua.edu.cn}}

\maketitle
\begin{abstract}
For uplink large-scale MIMO systems, linear minimum mean square error (MMSE) signal detection algorithm is near-optimal but involves matrix inversion with high complexity. In this paper, we propose a low-complexity signal detection algorithm based on the successive overrelaxation (SOR) method to avoid the complicated matrix inversion. We first prove a special property that the MMSE filtering matrix is symmetric positive definite for uplink large-scale MIMO systems, which is the premise for the SOR method. Then a low-complexity iterative signal detection algorithm based on the SOR method as well as the convergence proof is proposed. The analysis shows that the proposed scheme can reduce the computational complexity from ${{\cal O}({K^3})}$  to ${{\cal O}({K^2})}$, where ${K}$ is the number of users. Finally, we verify through simulation results that the proposed algorithm outperforms the recently proposed Neumann series approximation algorithm, and achieves the near-optimal performance of the classical MMSE algorithm with a small number of iterations.
\end{abstract}

\section{Introduction}\label{S1}

\IEEEPARstart Large-scale multiple-input multiple-output (MIMO) is considered as a key technology for future wireless communications~\cite{Dai12a}. Unlike the traditional small-scale MIMO technology (e.g., at most 8 antennas in LTE-A), large-scale MIMO exploits a very large number of antennas (e.g., 128 antennas or even more) at the base station (BS) to simultaneously serve multiple user equipments (UEs)~\cite{marzetta10}. It has been theoretically proved that large-scale MIMO can provide potential opportunity to increase the spectrum and energy efficiency by orders of magnitude~\cite{Dai13c}.

However, some challenging problems have to be solved to realize such attractive merits of large-scale MIMO in practice. One of them is the practical signal detection algorithm in the uplink~\cite{rusek13}. The optimal detector is the maximum likelihood (ML) detector whose complexity exponentially increases with the number of transmit antennas, which makes it impractical for large-scale MIMO systems. To achieve the (close) optimal ML detection performance with reduced complexity, several non-linear signal detection algorithms have been proposed. One typical category is based on the sphere decoding (SD) algorithm~\cite{wang13}, such as the fixed-complexity sphere decoding (FSD) algorithm~\cite{barbero08}. This kind of algorithms uses the underlying lattice structure of the received signal and considers the most promising approach to achieve the ML detection performance with reduced complexity. It performs well for the conventional small-scale MIMO systems, but when the dimension of the MIMO systems is large or the modulation order is high~\cite{Dai10b} (e.g., 128 antennas at the BS with 64 QAM modulation), the complexity is still unaffordable. Another category is based on the tabu search (TS) algorithm derived from artificial intelligence~\cite{datta10}, such as the layered tabu search (LTS) algorithm~\cite{srinidhi11}. This kind of algorithms utilizes the idea of local neighborhood search to estimate the transmitted signal and limits the selection of neighborhood by a tabu list. When the neighborhood range is appropriately small and the tabu list is carefully designed, the complexity is acceptable for large-scale MIMO systems, but it suffers from a non-negligible performance loss compared to the optimal ML detector. To make a trade-off between the performance and complexity, one can resort to linear signal detection algorithms, such as the zero-forcing (ZF) and minimum mean square error (MMSE) algorithms, which are near-optimal for uplink multi-user large-scale MIMO systems~\cite{rusek13}. However, these algorithms involve unfavorable inversion of a matrix of large size, whose complexity is still high for large-scale MIMO systems. Very recently, to reduce the complexity of matrix inversion, ~\cite{yin13} proposed the Neumann series approximation algorithm to convert the matrix inversion into a series of matrix-vector multiplications. However, only marginal reduction in complexity can be achieved.

In this paper, we propose a matrix inversion-less signal detection algorithm with low complexity based on the SOR method~\cite{bjorck1996numerical} for large-scale MIMO systems. We first prove that the MMSE filtering matrix is symmetric positive definite for uplink large-scale MIMO systems, according to which we propose to exploit the SOR method to avoid the complicated matrix inversion. We also prove the convergence of the proposed signal detection algorithm to guarantee its feasibility in practice. We verify through simulation results that the proposed algorithm can efficiently solve the matrix inversion problem in an iterative procedure until the desired detection accuracy is attained. To the best of our knowledge, this work is the first one to utilize the SOR method for the signal detection in uplink large-scale MIMO systems.

The rest of the paper is organized as follows. Section~\ref{S2} briefly describes the system model. Section~\ref{S3} specifies the proposed low-complexity signal detection algorithm, together with the convergence proof and the complexity analysis. The simulation results of the bit error rate (BER) performance are provided in Section~\ref{S4}. Finally, conclusions are drawn in Section~\ref{S5}.

{\it Notation}: We use lower-case and upper-case boldface letters to denote vectors and matrices, respectively; ${( \cdot )^T}$, ${( \cdot )^H}$, ${( \cdot )^{ - 1}}$, and $\left|  \cdot  \right|$ denote the transpose, conjugate transpose, matrix inversion, and absolute operators, respectively; ${{\mathop{\rm Re}\nolimits} \{  \cdot \} }$  and ${{\mathop{\rm Im}\nolimits} \{  \cdot \} }$ denote the real part and imaginary part of a complex number, respectively; Finally, ${{\bf{I}}_N}$ represents the $ N \times N $  identity matrix.

\section{System Model}\label{S2}
We consider a uplink large-scale MIMO system employing  ${N}$ antennas at the BS to simultaneously serve ${K}$ single-antenna
UEs~\cite{marzetta10, rusek13}. Usually we have ${N >  > K}$, e.g., ${N = 128}$ and ${K = 16}$
have been considered in~\cite{rusek13}.

The transmitted bit streams from different users are first encoded by the channel encoder and then mapped to symbols by taking values
from a modulation alphabet. Let ${{{\bf{s}}_c} = {[{s_{c,1}}, \cdot  \cdot  \cdot ,{s_{c,K}}]^T}}$ denote the
transmitted signal vector from all ${K}$  users, and ${{{\bf{H}}_c} \in {\mathbb{C}^{N \times K}}}$  denote the flat Rayleigh fading
channel matrix, whose entries are assumed to be independently and identically distributed (i.i.d.) with zero mean and unit variance~\cite{marzetta10}. Then the
received signal vector ${{{\bf{y}}_c} = {[{y_{c,1}}, \cdot  \cdot  \cdot ,{y_{c,N}}]^T}}$   at the BS can be represented as
\begin{equation}\label{eq1}
{{\bf{y}}_c}{\bf{ = }}{{\bf{H}}_c}{{\bf{s}}_c}{\bf{ + }}{{\bf{n}}_c},
\end{equation}
where ${{{\bf{n}}_c} = {[{n_{c,1}}, \cdot  \cdot  \cdot ,{n_{c,N}}]^T}}$ is the noise vector whose entries are i.i.d and follow the distribution ${{\cal CN}(0,{\sigma ^2})}$.

%

For signal detection, the complex-valued system model (1) can be converted into a corresponding real-valued one as
\begin{equation}\label{eq2}
{\bf{y = Hs + n}},
\end{equation}
where ${{\bf{y}} = {[{\mathop{\rm Re}\nolimits} \{ {{\bf{y}}_c}\} \quad {\mathop{\rm Im}\nolimits} \{ {{\bf{y}}_c}\} ]^T}}$ is of size ${2N \times 1}$, accordingly ${{\bf{s}} = {[{\mathop{\rm Re}\nolimits} \{ {{\bf{s}}_c}\} \quad {\mathop{\rm Im}\nolimits} \{ {{\bf{s}}_c}\} ]^T}}$, ${{\bf{n}} = {[{\mathop{\rm Re}\nolimits} \{ {{\bf{n}}_c}\} \quad {\mathop{\rm Im}\nolimits} \{ {{\bf{n}}_c}\} ]^T}}$, and
\begin{equation}\label{eq3}
{\bf{H}} = {\left[ \begin{array}{l}
{\mathop{\rm Re}\nolimits} \{ {{\bf{H}}_c}\} \quad  - {\mathop{\rm Im}\nolimits} \{ {{\bf{H}}_c}\} \\
{\mathop{\rm Im}\nolimits} \{ {{\bf{H}}_c}\} \quad \,\;\,{\mathop{\rm Re}\nolimits} \{ {{\bf{H}}_c}\}
\end{array} \right]_{2N \times 2K}}.
\end{equation}

At the BS, after the channel matrix ${{\bf{{H}}}}$ has been obtained through time-domain and/or frequency-domain training pilots~\cite{Dai12b}~\cite{Erik13},
the task of signal detection is to
recover the transmitted signal vector ${{\bf{s}}}$  from the received signal vector ${{\bf{y}}}$. It has been proved that the linear MMSE signal detection
algorithm is near-optimal for uplink multi-user large-scale MIMO systems~\cite{rusek13}, and the estimate of the transmitted signal vector ${{\bf{\hat s}}}$
can be obtained by
\begin{equation}\label{eq4}
{\bf{\hat s}} = {({{\bf{H}}^H}{\bf{H}} + {\sigma ^2}{{\bf{I}}_{2K}})^{ - 1}}{{\bf{H}}^H}{\bf{y}} = {{\bf{W}}^{ - 1}}\hat {\bf{y}},
\end{equation}
where ${{\bf{\hat y}} = {{\bf{H}}^H}{\bf{y}}}$, and the MMSE filtering matrix ${{\bf{{W}}}}$ is denoted as
\begin{equation}\label{eq5}
{\bf{W}} = {\bf{G}} + {\sigma ^2}{{\bf{{I}}}_{2K}},
\end{equation}
where ${{\bf{G}} = {{\bf{{H}}}^H}{\bf{{H}}}}$ is the Gram matrix. The computational complexity of the direct matrix inversion ${{{\bf{W}}^{ - 1}}}$ is ${{\cal O}({K^3})}$,
which is high for large-scale MIMO systems.

\section{Low-Complexity Signal Detection For Uplink Large-Scale MIMO}\label{S3}
In this section, we first prove a special property of large-scale MIMO systems that the MMSE filtering matrix is
symmetric positive definite. Based on this property, we then propose a low-complexity signal detection algorithm utilizing the SOR method to iteratively
achieve the MMSE estimate without matrix inversion. The convergence proof is also addressed. Finally, we provide the complexity analysis of the proposed algorithm
to show its advantage over conventional schemes.

\subsection{Matrix inversion-less signal detection utilizing SOR method}\label{S2.1}
Unlike the conventional (small-scale) MIMO systems with small number of antennas, large-scale MIMO systems enjoy a special property that the column vectors of the
channel matrix are asymptotically orthogonal~\cite{rusek13}. Based on that, we prove that the MMSE filtering matrix is symmetric positive definite in the following \textbf{Lemma 1}.

\vspace*{+2mm} \noindent\textbf{Lemma 1.} {\it For uplink large-scale MIMO systems, the MMSE filtering matrix ${{\bf{W}}}$  is symmetric positive definite}.
\vspace*{+2mm}

\textit{Proof:} Since the complex-valued MIMO system model has been converted  into the real-valued one, the transpose of matrix and the conjugate transpose of matrix will be the
same, e.g., ${{\bf{G}} = {{\bf{H}}^H}{\bf{H}} = {{\bf{H}}^T}{\bf{H}}}$. Thus, we have
\begin{equation}\label{eq6}
{{\bf{G}}^T} = {({{\bf{H}}^T}{\bf{H}})^T} = {{\bf{H}}^T}{\bf{H}} = {\bf{G}},
\end{equation}
which indicates that the Gram matrix ${{\bf{G}}}$ is symmetric. Meanwhile, for uplink large-scale MIMO systems, the column vectors of the real-valued channel
matrix ${{\bf{H}}}$ are asymptotically orthogonal~\cite{rusek13}, i.e., the equation ${{\bf{Hq}} = 0}$  has an unique solution, which is the ${2K \times 1}$
zero vector. Thus, for any ${2K \times 1}$ non-zero real-valued vector  ${{\bf{r}}}$, we have
\begin{equation}\label{eq7}
{\left( {{\bf{Hr}}} \right)^T}{\bf{Hr}} = {{\bf{r}}^T}{\bf{Gr}} > 0,
\end{equation}
which implies that ${{\bf{G}}}$  is positive definite. Considering (6) and (7), we can conclude that the Gram matrix ${{\bf{G = }}{{\bf{H}}^T}{\bf{H}}}$ is
symmetric positive definite. Finally, as the noise variance ${{\sigma ^2}}$ is positive, the MMSE filtering
matrix ${{\bf{W}} = {\bf{G}} + {\sigma ^2}{{\bf{I}}_{2K}}}$  in (5) is symmetric positive definite, too.   \qed

The special property that the MMSE filtering matrix ${{\bf{W}}}$ in uplink large-scale MIMO systems is symmetric positive definite inspires us to exploit the
SOR method to
efficiently solve (4) with low complexity. The SOR method is used to solve ${N}$-dimension linear equation ${{\bf{Ax}} = {\bf{b}}}$, where ${{\bf{A}}}$ is the
${N \times N}$ symmetric positive definite matrix, ${{\bf{x}}}$
is the ${N \times 1}$ solution vector, and ${{\bf{b}}}$ is the ${N \times 1}$  measurement vector. Unlike the traditional method that directly computes ${{{\bf{A}}^{ - 1}}{\bf{b}}}$
 to obtain ${{\bf{x}}}$, the SOR method can efficiently solve
the linear equation in an iterative manner without the complicated matrix inversion.
Since matrix ${{\bf{A}}}$ is symmetric positive definite, we can decompose it into a diagonal component ${{{\bf{D}}_{\bf{A}}}}$, a strictly lower triangular
component ${{{\bf{L}}_{\bf{A}}}}$,
and a strictly upper triangular component ${{\bf{L}}_{\bf{A}}^T}$. Then the SOR iteration can be described as~\cite{bjorck1996numerical}
\begin{equation}\label{eq8}
\begin{split}
&{{\bf{x}}^{(i + 1)}} \\
&={({{\bf{L}}_{\bf{A}}}\! +\! \frac{1}{w}{{\bf{D}}_{\bf{A}}})^{ - 1}}\left[ {\left( {(\frac{1}{w}\! -\! 1){{\bf{D}}_{\bf{A}}}\! -\! {\bf{L}}_{\bf{A}}^T} \right){{\bf{x}}^{(i)}}\! +\! {\bf{b}}} \right],
\end{split}
\end{equation}
where the superscript ${i = 0,1,2, \cdot  \cdot  \cdot}$ denotes the number of iterations, and ${w}$ represents the relaxation parameter, which plays an important role in the convergence and the convergence rate.
Note that when ${w = 1}$, the SOR method is the same as the well known Gauss-Seidel method~\cite{bjorck1996numerical}, which means that the Gauss-Seidel method is a special case of the SOR method.
We will discuss the selection of the relaxation parameter ${w}$ in detail later in Section~\ref{S4}.

Due to the MMSE filtering matrix ${{\bf{W}}}$ is symmetric positive definite for uplink large-scale MIMO systems as proved in \textbf{Lemma 1},
we can also decompose ${{\bf{W}}}$  as
\begin{equation}\label{eq9}
{\bf{W}} = {\bf{D}} + {\bf{L}} + {{\bf{L}}^T},
\end{equation}
where ${{\bf{D}}}$, ${{\bf{L}}}$, and ${{{\bf{L}}^T}}$ denote the diagonal component, the strictly lower triangular component, and the strictly upper triangular component of ${{\bf{W}}}$, respectively.
Then we can utilize the SOR method to estimate the transmitted signal vector ${{\bf{s}}}$ as below
\begin{equation}\label{eq10}
{{\bf{s}}^{(i + 1)}}\! =\! {({\bf{L}}\! +\! \frac{1}{w}{\bf{D}})^{ - 1}}\left[ {\left( {(\frac{1}{w}\! -\! 1){\bf{D}}\! -\! {{\bf{L}}^T}} \right){{\bf{s}}^{(i)}}\! +\! {\bf{\hat y}}} \right],
\end{equation}
where ${{{\bf{s}}^{(0)}}}$ denotes the initial solution, which is usually set as a ${2K \times 1}$  zero vector without loss of generality~\cite{bjorck1996numerical}.
Then the solution to the signal detection problem (4) can be solved by the SOR method according to
\begin{equation}\label{eq11}
({\bf{L + }}\frac{1}{w}{\bf{D}}){{\bf{s}}^{(i + 1)}} = {\bf{\hat y}} + \left( {(\frac{1}{w} - 1){\bf{D}} - {{\bf{L}}^T}} \right){{\bf{s}}^{(i)}}.
\end{equation}
As ${({\bf{L + }}\frac{1}{w}{\bf{D}})}$ is a lower triangular matrix, one can solve the equation (11) to obtain ${{{\bf{s}}^{(i + 1)}}}$  with low complexity
as will be addressed in Section~\ref{S3}-C.
Next we will prove the convergence of the proposed signal detection based on the SOR method.

\subsection{Convergence proof}\label{S2.2}
\vspace*{+2mm} \noindent\textbf{Lemma 2.} {\it For uplink large-scale MIMO systems, the signal detection algorithm using the SOR method is convergent when the relaxation parameter ${w}$ satisfies ${0 < w < 2}$}.
\vspace*{+2mm}

\textit{Proof:} We define ${{\bf{C}} = {({\bf{L}} + \frac{1}{w}{\bf{D}})^{ - 1}}(\frac{1}{w}{\bf{D}} - {\bf{D}} - {{\bf{L}}^T})}$
and ${{\bf{d}} = {({\bf{L}} + \frac{1}{w}{\bf{D}})^{ - 1}}{\bf{\hat y}}}$, where ${{\bf{C}}}$ is called as the iteration matrix. Then the SOR iteration (11) can be rewritten as
\begin{equation}\label{eq12}
{{\bf{s}}^{(i + 1)}} = {\bf{C}}{{\bf{s}}^{(i)}} + {\bf{d}}.
\end{equation}

The spectral radius of the iteration matrix ${{\bf{C}}}$  is defined as the non-negative number  ${\rho ({\bf{C}}) = \mathop {\max }\limits_{1 \le n \le 2K} \left| {{\lambda _n}} \right|}$,
where ${{\lambda _n}}$  denotes the  ${n}$th eigenvalue of ${{\bf{C}}}$.
The necessary and sufficient conditions for the convergence of (12) is that the spectral radius should satisfy~\cite[Theorem 7.2.2]{bjorck1996numerical}
\begin{equation}\label{eq13}
\rho ({\bf{C}}) = \mathop {\max }\limits_{1 \le n \le 2K} \left| {{\lambda _n}} \right| < 1.
\end{equation}
According to the definition of eigenvalue, we have
\begin{equation}\label{eq14}
{\bf{Cr}} = {({\bf{L}} + \frac{1}{w}{\bf{D}})^{ - 1}}(\frac{1}{w}{\bf{D}} - {\bf{D}} - {{\bf{L}}^T}){\bf{r}} = {\lambda _n}{\bf{r}},
\end{equation}
where ${{\bf{r}}}$ is an arbitrary ${2K \times 1}$ non-zero real-valued vector. Note that (14) can be also presented as
\begin{equation}\label{eq15}
(\frac{1}{w}{\bf{D}} - {\bf{D}} - {{\bf{L}}^T}){\bf{r}} = ({\bf{L}} + \frac{1}{w}{\bf{D}}){\lambda _n}{\bf{r}}.
\end{equation}
Multiply both sides of (15) by ${{{\bf{r}}^T}}$ will yield
\begin{equation}\label{eq16}
{{\bf{r}}^T}(\frac{1}{w}{\bf{D}} - {\bf{D}} - {{\bf{L}}^T}){\bf{r}} = {\lambda _n}{{\bf{r}}^T}({\bf{L}} + \frac{1}{w}{\bf{D}}){\bf{r}}.
\end{equation}
Then we take transpose on both sides of (16), and another equation can be obtained as
\begin{equation}\label{eq17}
{{\bf{r}}^T}(\frac{1}{w}{\bf{D}} - {\bf{D}} - {\bf{L}}){\bf{r}} = {\lambda _n}{{\bf{r}}^T}({{\bf{L}}^T} + \frac{1}{w}{\bf{D}}){\bf{r}}.
\end{equation}
Note that ${{\bf{D}} = {{\bf{D}}^T}}$ as ${{\bf{D}}}$ is a diagonal matrix. Add (16) and (17) will lead to
\begin{equation}\label{eq18}
{{\bf{r}}^T}\left( {(\frac{2}{w} - 2){\bf{D}} - {\bf{L}} - {{\bf{L}}^T}} \right){\bf{r}}\! =\! {\lambda _n}{{\bf{r}}^T}({{\bf{L}}^T} + {\bf{L}} + \frac{2}{w}{\bf{D}}){\bf{r}}.
\end{equation}
Substituting (9) into (18), we have
\begin{equation}\label{eq19}
(1 - {\lambda _n})(\frac{2}{w} - 1){{\bf{r}}^T}{\bf{Dr}} = (1 + {\lambda _n}){{\bf{r}}^T}{\bf{Wr}}.
\end{equation}
Since the MMSE filtering matrix ${{\bf{W}}}$ is positive definite as proved above, the diagonal matrix ${{\bf{D}}}$ is positive definite, too. Then we have ${{{\bf{r}}^T}{\bf{Dr}} > 0}$ and ${{{\bf{r}}^T}{\bf{Wr}} > 0}$. Besides, we also have ${(\frac{2}{w} - 1) > 0}$ if ${0 < w < 2}$. Thus, we can conclude that ${(1 - {\lambda _n})(1 + {\lambda _n}) > 0}$, which means
\begin{equation}\label{eq20}
\left| {{\lambda _n}} \right| < 1.
\end{equation}
Substituting (20) into (13), we can assert that ${\rho ({\bf{C}}) < 1}$, so the SOR iteration (11) is convergent. \qed

It is worth pointing out that another different proof of \textbf{Lemma 2}
can be found in~\cite[Theorem 11.2.3]{golub2012matrix}, which utilizes the
orthogonal transformation  with high complexity to obtain the convergence proof, while
our method directly exploits the definition of eigenvalue, which is simpler than the existing method~\cite{golub2012matrix}.

\subsection{Computational complexity analysis}\label{S2.3}
The computational complexity in terms of required number of multiplications is analyzed in this part.
It can be found from (11) that the computational complexity of the ${i}$th iteration of the proposed signal detection
algorithm originates from solving the linear equation.
Considering the definition of ${{\bf{D}}}$, ${{\bf{L}}}$, and  ${{{\bf{L}}^T}}$, the solution can be presented as
\begin{equation}\label{eq21}
\begin{split}
&s_m^{(i + 1)} = (1 - w)s_m^{(i)} \\
&+ \frac{w}{{{W_{m,m}}}}({\hat y_m} - \sum\limits_{k < m} {{W_{m,k}}s_k^{(i + 1)} - \sum\limits_{k > m} {{W_{m,k}}s_k^{(i)}} } ),\\
&\qquad \qquad \qquad  m,k = 1,2, \cdot  \cdot  \cdot 2K,
\end{split}
\end{equation}
where ${s_m^{(i)}}$, ${s_m^{(i + 1)}}$, and ${{\hat y_m}}$ denote the ${m}$th element of ${{{\bf{s}}^{(i)}}}$, ${{{\bf{s}}^{(i+1)}}}$, and ${{\bf{\hat y}}}$  in
(4), respectively,
and ${{W_{m,k}}}$ denotes the  ${m}$th row and  ${k}$th column entry of ${{\bf{W}}}$.
The required number of multiplications in the computation of ${(1 - w)s_m^{(i)}}$  and ${\frac{w}{{{W_{m,m}}}}({\hat y_m} - \sum\limits_{k < m} {{W_{m,k}}s_k^{(i + 1)} - \sum\limits_{k > m} {{W_{m,k}}s_k^{(i)}} } )}$
is 1 and ${2K + 1}$, respectively.
Therefore the computation of each element of ${{{\bf{s}}^{(i + 1)}}}$ requires ${2K + 2}$  times of multiplications. Since there are ${2K}$ elements in ${{{\bf{s}}^{(i + 1)}}}$,
the overall required number of multiplications is ${4{K^2} + 4K}$.

\begin{table}[h]
\setlength{\abovecaptionskip}{-10pt}
\setlength{\belowcaptionskip}{0pt}
\caption{Computational Complexity} \label{TAB1}
\begin{center}
\begin{threeparttable}
\begin{tabular}{*{1}{L{0.5cm}}*{2}{C{3.5cm}}}
\toprule[1pt]
 & Conventional Neumann series approximation algorithm~\cite{yin13} & Proposed signal detection algorithm\\
\hline \\ [-2 ex]
 ${i = 2}$ & ${12{K^2} - 4K}$ & ${8{K^2} + 8K}$\\
 ${i = 3}$ & ${8{K^3} + 4{K^2} - 2K}$  & ${12{K^2} + 12K}$\\
 ${i = 4}$ & ${16{K^3} - 4{K^2}}$  & ${16{K^2} + 16K}$\\
 ${i = 5}$ & ${24{K^3} - 12{K^2} + 2K}$  & ${20{K^2} + 20K}$\\
\toprule[1pt]
\end{tabular}
\end{threeparttable}
\end{center}
\end{table}
\vspace*{-4mm}

Table~\ref{S1} compares the complexity of the conventional Neumann series approximation algorithm~\cite{yin13} and the
proposed algorithm based on the SOR method. Since the complexity of the classical MMSE algorithm is ${{\cal O}({K^3})}$,
we can conclude from Table~\ref{S1} that the conventional Neumann series approximation algorithm can reduce the
complexity from ${{\cal O}({K^3})}$  to ${{\cal O}({K^2})}$ when the number of iterations is  ${i = 2}$, but the complexity is still ${{\cal O}({K^3})}$  when ${i \ge 3}$.
To ensure the approximation performance, usually a large value of ${i}$ is required to approach the final
MMSE solution  ${{\bf{\hat s}}}$ as will be verified later in Section~\ref{S4}. So the overall complexity is almost the
same as the MMSE algorithm, which means only marginal reduction in complexity can be achieved.
However, we can observe that the complexity of the proposed algorithm is ${{\cal O}({K^2})}$
for arbitrary number of iterations. And even for ${i = 2}$, the proposed algorithm enjoys a
lower complexity than the conventional one~\cite{yin13}.

Additionally, we can observe from (21) that the computation of ${s_m^{(i + 1)}}$ utilizes ${s_k^{(i + 1)}}$ for ${k = 1,2, \cdot  \cdot  \cdot ,m - 1}$ and ${s_l^{(i)}}$ for ${l = m,m + 1, \cdot  \cdot  \cdot ,2K}$,
which is similar to the Gauss-Seidel method~\cite{bjorck1996numerical}.
Then, two another benefits can be expected. Firstly, after ${s_m^{(i + 1)}}$  has been obtained, we can use it to overwrite ${s_m^{(i)}}$ which is useless in the next computation of  ${s_{m + 1}^{(i + 1)}}$.
Consequently, only one storage vector of size ${2K \times 1}$ is required; secondly, when ${i}$ increases, the solution to (11) becomes closer to the final MMSE
solution  ${{\bf{\hat s}}}$. Thus ${s_m^{(i + 1)}}$ can exploits the elements of ${s_k^{(i + 1)}}$ for ${k = 1,2, \cdot  \cdot  \cdot ,m - 1}$  that have already
been  computed in the current iteration to produce more reliable result than the conventional algorithm~\cite{yin13} only utilizing all the elements of
${{{\bf{s}}^{(i)}}}$ in the previous iteration. Thus, a faster convergence rate can be expected, and the required number of iterations to achieve a certain
estimation accuracy becomes smaller.
Based on these two special advantages of the SOR method, the overall complexity of the proposed algorithm can be reduced further.

\section{Simulation Results}\label{S4}
To verify the performance of the proposed signal detection algorithm, we provide the BER simulation results compared with the recently proposed
Neumann series approximation algorithm~\cite{yin13}. The BER performance of the classical MMSE algorithm with complicated but exact matrix inversion
is included as the benchmark for comparison. Besides, to verify the near-optimal performance of the MMSE algorithm, the performance of the optimal ML detection algorithm  is also provided. We consider two large-scale MIMO systems with ${N \times K = 64 \times 8}$  and ${N \times K = 128 \times 16}$,
respectively. The modulation scheme
of 64 QAM is adopted.  The rate-1/2 industry standard convolutional code with generator polynomials ${[{133_o}\;{171_o}]}$ is employed, and a random interleaver is
also used to combat the burst error. The Rayleigh fading channel model is considered. After multi-user signal detection, the estimated signal vector is used to extract
the soft information (by calculate the log-likelihood ratios (LLRs)) for soft-input Viterbi decoder for channel decoding.

\begin{figure}[tp]
\setlength{\abovecaptionskip}{-10pt}
\setlength{\belowcaptionskip}{0pt}
\begin{center}
\vspace*{-6mm}\includegraphics[width=0.93\linewidth]{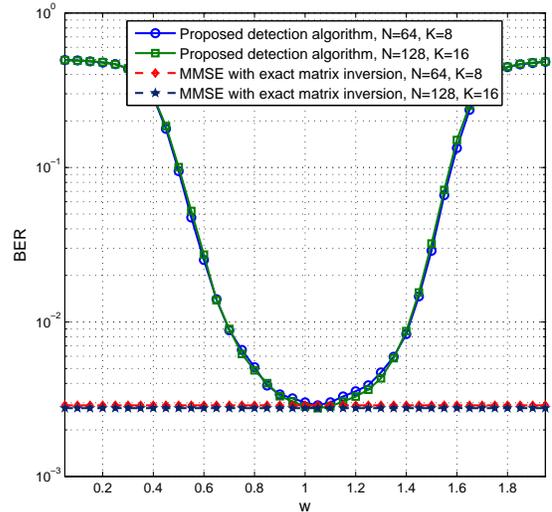}
\end{center}
\caption{BER performance of the proposed SOR-based signal detection algorithm against the relaxation parameter  ${w}$, where SNR = 4 dB and ${i = 3}$.} \label{FIG3}
\end{figure}

\begin{figure}[tp]
\setlength{\abovecaptionskip}{-10pt}
\setlength{\belowcaptionskip}{0pt}
\begin{center}
\vspace*{-5mm}\includegraphics[width=0.93\linewidth]{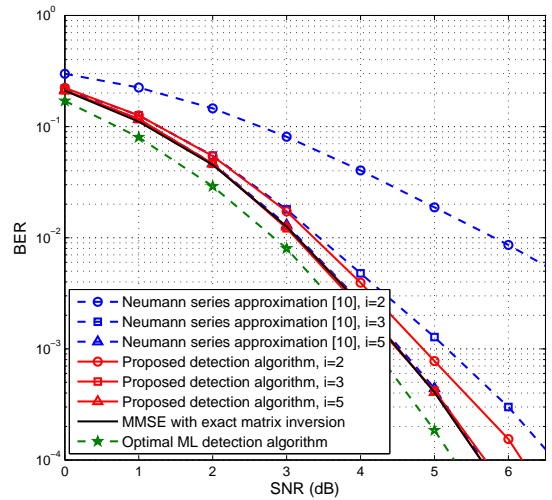}
\end{center}
\caption{BER performance comparison when  ${N \times K = 64 \times 8}$.} \label{FIG3}
\end{figure}

\begin{figure}[tp]
\setlength{\abovecaptionskip}{-10pt}
\setlength{\belowcaptionskip}{0pt}
\begin{center}
\vspace*{-6mm}\includegraphics[width=0.93\linewidth]{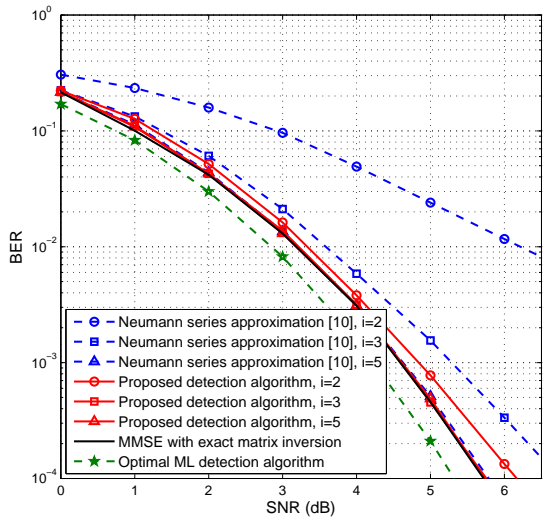}
\end{center}
\caption{BER performance comparison when  ${N \times K = 128 \times 16}$.} \label{FIG3}
\end{figure}

Fig. 1 shows the BER performance of the proposed SOR-based signal detection algorithm against the relaxation parameter ${w}$,
where the signal-to-noise ratio (SNR) is 4 dB, and the number of iterations is ${i = 3}$. As shown in Fig. 1, BERs of the MMSE algorithm
are ${2.8854 \times {10^{ - 3}}}$  for ${N \times K = 64 \times 8}$,  and ${2.7656 \times {10^{ - 3}}}$  for ${N \times K = 128 \times 16}$, respectively,
which are the targets to be approached by selecting the optimal relaxation parameters.
We can observe that the BER curve against ${w}$ looks like a parabola, and fortunately the optimal ${w}$  for both systems
is 1.05. Furthermore, we have conducted intensive simulations of different large-scale MIMO system configurations and
found that the systems with fixed ${N/K}$ (e.g., ${N/K = 8}$ in Fig. 1) will share the same optimal selection of ${w}$,
which indicates that we can easily obtain the optimal ${w}$ after the system dimensions ${N}$  and ${K}$ have been fixed.

The BER performance comparison between the conventional Neumann series approximation algorithm~\cite{yin13} and the proposed SOR-based
signal detection algorithm when ${N \times K = 64 \times 8}$ and  ${N \times K = 128 \times 16}$ are shown in Fig. 2 and Fig. 3, respectively,
where ${i}$ denotes the number of iterations.
It is clear that the BER performance of both algorithms improves with the increased number of iterations. However, when the
same iteration number ${i}$ is used, the proposed algorithm outperforms the conventional one for both systems. Moreover, as we can
observe from Fig. 2, the BER performance of the proposed algorithm when ${i = 3}$ is almost the same as that of the conventional one when ${i = 5}$,
which indicates that a faster convergence rate can be achieved by the proposed SOR-based signal detection algorithm.
As we have addressed in Section~\ref{S3}-C, a faster convergence rate means smaller number of iterations is required to achieve a certain
estimation accuracy, so the complexity of the proposed algorithm can be reduced further.

Meanwhile, we can observe from Fig. 2 and Fig. 3 that the MMSE algorithm is near-optimal compared to the optimal ML detection algorithm, and the
proposed algorithm without the complicated matrix inversion can achieve the near-optimal BER performance of the MMSE algorithm when the number of iterations is large (e.g., ${i = 3}$ in Fig. 2 and Fig. 3).

\section{Conclusions}\label{S5}
In this paper, by fully exploiting a special channel property of the large-scale MIMO systems, we
propose a low-complexity near-optimal signal detection algorithm based on the SOR method in the uplink.
The SOR-based algorithm can iteratively realize the MMSE solution without complicated matrix inversion,
which can reduce the complexity from ${{\cal O}({K^3})}$ to ${{\cal O}({K^2})}$. We also prove the convergence of the proposed algorithm,
and simulation results show that it can achieve the near-optimal performance of the classical MMSE algorithm
with a small number of iterations. Moreover, the idea of utilizing the SOR method to efficiently realize matrix
inversion with low complexity can be extended to other signal processing problems in wireless communications,
such as the precoding in the  large-scale MIMO systems.

\vspace*{0mm}
\section*{Acknowledgments}
This work was supported by National Key Basic Research Program of
China (Grant No. 2013CB329201), National Natural Science Foundation
of China (Grant Nos. 61271266 and 61201185),  Science and Technology Foundation for Beijing Outstanding Doctoral Dissertation (Grant No. 2012T50093), and the ZTE fund project CON1307250001.

\vspace*{0mm}
\bibliography{IEEEabrv,Gao1Ref}

\begin{thebibliography}{10}
\providecommand{\url}[1]{#1}
\csname url@samestyle\endcsname
\providecommand{\newblock}{\relax}
\providecommand{\bibinfo}[2]{#2}
\providecommand{\BIBentrySTDinterwordspacing}{\spaceskip=0pt\relax}
\providecommand{\BIBentryALTinterwordstretchfactor}{4}
\providecommand{\BIBentryALTinterwordspacing}{\spaceskip=\fontdimen2\font plus
\BIBentryALTinterwordstretchfactor\fontdimen3\font minus
  \fontdimen4\font\relax}
\providecommand{\BIBforeignlanguage}[2]{{%
\expandafter\ifx\csname l@#1\endcsname\relax
\typeout{** WARNING: IEEEtran.bst: No hyphenation pattern has been}%
\typeout{** loaded for the language `#1'. Using the pattern for}%
\typeout{** the default language instead.}%
\else
\language=\csname l@#1\endcsname
\fi
#2}}
\providecommand{\BIBdecl}{\relax}
\BIBdecl

\bibitem{Dai12a}
L.~Dai, Z.~Wang, and Z.~Yang, ``Next-generation digital television terrestrial
  broadcasting systems: {K}ey technologies and research trends,'' \emph{{IEEE}
  Commun. Mag.}, vol.~50, no.~6, pp. 150--158, Jun. 2012.

\bibitem{marzetta10}
T.~L. Marzetta, ``Noncooperative cellular wireless with unlimited numbers of
  base station antennas,'' \emph{{IEEE} Trans. Wireless Commun.}, vol.~9,
  no.~11, pp. 3590--3600, Nov. 2010.

\bibitem{Dai13c}
L.~Dai, Z.~Wang, and Z.~Yang, ``Spectrally efficient time-frequency training
  {OFDM} for mobile large-scale {MIMO} systems,'' \emph{{IEEE} J. Sel. Areas
  Commun.}, vol.~31, no.~2, pp. 251--263, Feb. 2013.

\bibitem{rusek13}
F.~Rusek, D.~Persson, B.~K. Lau, E.~G. Larsson, T.~L. Marzetta, O.~Edfors, and
  F.~Tufvesson, ``Scaling up {MIMO}: Opportunities and challenges with very
  large arrays,'' \emph{{IEEE} Signal Process. Mag.}, vol.~30, no.~1, pp.
  40--60, Jan. 2013.

\bibitem{wang13}
Y.~Wang and H.~Leib, ``Sphere decoding for {MIMO} systems with {N}ewton
  iterative matrix inversion,'' \emph{{IEEE} Commun. Lett.}, vol.~17, no.~2,
  pp. 389--392, Feb. 2013.

\bibitem{barbero08}
L.~G. Barbero and J.~S. Thompson, ``Fixing the complexity of the sphere decoder
  for {MIMO} detection,'' \emph{{IEEE} Trans. Wireless Commun.}, vol.~7, no.~6,
  pp. 2131--2142, Jun. 2008.

\bibitem{Dai10b}
R.~Ma, L.~Dai, Z.~Wang, and J.~Wang, ``Secure communication in {TDS-OFDM}
  system using constellation rotation and noise insertion,'' \emph{{IEEE}
  Trans. Consum. Electron.}, vol.~56, no.~3, pp. 1328--1332, Aug. 2010.

\bibitem{datta10}
T.~Datta, N.~Srinidhi, A.~Chockalingam, and B.~S. Rajan, ``Random-restart
  reactive tabu search algorithm for detection in large-{MIMO} systems,''
  \emph{{IEEE} Commun. Lett.}, vol.~14, no.~12, pp. 1107--1109, Dec. 2010.

\bibitem{srinidhi11}
N.~Srinidhi, T.~Datta, A.~Chockalingam, and B.~S. Rajan, ``Layered tabu search
  algorithm for large-{MIMO} detection and a lower bound on {ML} performance,''
  \emph{{IEEE} Trans. Commun.}, vol.~59, no.~11, pp. 2955--2963, Nov. 2011.

\bibitem{yin13}
B.~Yin, M.~Wu, C.~Studer, J.~R. Cavallaro, and C.~Dick, ``Implementation
  trade-offs for linear detection in large-scale {MIMO} systems,'' in
  \emph{Proc. {IEEE} International Conference on Acoustics, Speech and Signal
  Processing (ICASSP'13)}, May 2013, pp. 2679--2683.

\bibitem{bjorck1996numerical}
A.~Bj{\"o}rck, \emph{Numerical Methods for Least Squares Problems}.\hskip 1em
  plus 0.5em minus 0.4em\relax Society for Industrial and Applied Mathematics
  (SIAM), 1996.

\bibitem{Dai12b}
L.~Dai, Z.~Wang, and Z.~Yang, ``Time-frequency training {OFDM} with high
  spectral efficiency and reliable performance in high speed environments,''
  \emph{{IEEE} J. Sel. Areas Commun.}, vol.~30, no.~4, pp. 695--707, May 2012.

\bibitem{Erik13}
H.~A. Suraweera, H.~Q. Ngo, T.~Q. Duong, C.~Yuen, and E.~G. Larsson,
  ``Multi-pair amplify- and -foward relaying with very large antena array,'' in
  \emph{Proc. {IEEE} International Conference on Communication (ICC'13)}, Jun.
  2013, pp. 4635--4640.

\bibitem{golub2012matrix}
G.~H. Golub and C.~F. Van~Loan, \emph{Matrix computations}.\hskip 1em plus
  0.5em minus 0.4em\relax JHU Press, 2012.

\end{thebibliography}

\balance

\end{document}